\documentclass[aps,prl,showpacs,twocolumn,floats]{revtex4}
\usepackage{amssymb}

\usepackage{graphicx,psfig}
\usepackage{dcolumn}
\usepackage{bm}

\begin{document}

\bibliographystyle{prsty}

\title{
Rabi oscillations from ultrasound in spin systems \vspace{-1mm} }
\author{
C. Calero and E. M. Chudnovsky}
 \affiliation{\mbox{Department of Physics and Astronomy,} \\
\mbox{Lehman College, City University of New York,} \\ \mbox{250
Bedford Park Boulevard West, Bronx, New York 10468-1589, U.S.A.}
\\ {\today}}
\begin{abstract}
It is shown that ultrasound in the GHz range can generate
space-time Rabi oscillations between spin states of molecular
magnets. We compute dynamics of the magnetization generated by
surface acoustic waves and discuss conditions under which this
novel quantum effect can be observed.
\end{abstract}


\maketitle

Quantum mechanics of a spin cluster (e.g., a magnetic molecule)
embedded in a solid is determined by the crystal field. The latter
depends on the symmetry of the cluster and its environment
\cite{book}. Crystal field Hamiltonians provide good description
of molecular magnets at low temperature. For instance, dynamics of
a spin ${\bf S}$ that prefers to look up or down along the
anisotropy axis of the cluster can be described by a Hamiltonian
${\cal{H}} = -DS_z^2$, where the anisotropy constant $D$ arises
from spin-orbit interactions. We are interested in the effect of
the mechanical rotation of the crystal on the molecular spin. For
a similar problem involving an orbital moment, ${\bf L}$, it is
well known from classical mechanics that in the reference frame
rotating at an angular velocity ${\bf \Omega}$ the Hamiltonian
acquires a term $- \hbar {\bf L}\cdot {\bf \Omega} $ (we use
dimensionless $L$ and $S$). The same rule is expressed by the
Larmor theorem in classical electrodynamics: Rotation is
equivalent to the magnetic field, leading to the effective Zeeman
term, $- {\bf M}\cdot {\bf B}$, in the rotating-frame Hamiltonian,
with ${\bf M} = \hbar \gamma {\bf L}$, ${\bf B} = {\bf
\Omega}/\gamma$, and $\gamma$ being the gyromagnetic ratio. The
extension of the Larmor theorem to a spin is a consequence of the
fact that in relativistic quantum theory the generator of
rotations is $\hat{\bf J} = \hat{\bf L} + \hat{\bf S}$. Rigorous
derivation of the term $-(\hat{\bf L} + \hat{\bf S})\cdot {\bf
\Omega}$ in the Hamiltonian can be obtained, e.g., from the study
of the non-relativistic limit of the Dirac equation in the
rotating frame \cite{Dirac}.

Equivalence of the rotation to the magnetic field explains Barnett
effect \cite{Barnett}: Rotation of a body of the magnetic
susceptibility $\chi$ at an angular velocity ${\bf \Omega}$
generates a magnetic moment ${\bf M} = \chi {\bf \Omega}/\gamma$.
The ``spin-rotation coupling'', $-\hbar{\bf S}\cdot {\bf \Omega}$,
can also lead to non-trivial quantum effects. Consider, e.g., a
spin cluster with the Hamiltonian $\hat{\cal{H}} = -D\hat{S}_z^2$
that preserves the direction of the spin along the anisotropy axis
$Z$ due to commutation of $\hat{\cal{H}}$ with $\hat{S}_z$. In the
presence of the rotation about, e.g., the X-axis of the crystal
the Hamiltonian in the rotating frame becomes $\hat{\cal{H}}' =
-D\hat{S}_z^2 - \hbar \hat{S}_x\Omega$. This Hamiltonian, unlike
$\hat{\cal{H}}$, does not commute with $\hat{S}_z$ and, therefore,
allows transitions between the two orientations of ${\bf S}$ along
the anisotropy axis. Thus, rotation alone can induce quantum
transitions between spin states that are prohibited by the
Hamiltonian of a stationary system $\hat{\cal{H}}$. We should
emphasize that switching from the laboratory-frame Hamiltonian,
$\hat{\cal{H}}$, to the rotating-frame Hamiltonian,
$\hat{\cal{H}}' = \hat{\cal{H}} - \hbar \hat{\bf S}\cdot {\bf
\Omega}$, does not introduce any new spin-lattice interactions in
addition to the crystal field. It is just another method to obtain
solution of the problem, which, in the laboratory frame, requires
introduction of the time dependence of the crystal field: e.g.,
$\hat{\cal{H}}=-D\hat{S}_z^2$, in the presence of rotation,
becomes $\hat{\cal{H}}=-D[{\bf n}(t)\cdot\hat{\bf S}]^2$ with
${\bf n}(t)$ being the instantaneous direction of the anisotropy
axis.

So far, quantum spin-rotation effects received little attention
because only a very tiny magnetic field due to rotation can be
produced in the rotating frame of a macroscopic body.
Consequently, the corresponding quantum effects have very low
probability. This Letter is based upon the observation that {\it
local} rotations of the crystal lattice produced by high-frequency
ultrasound can easily provide $10\,$G - $100\,$G fields in the
rotating frame of a rigid spin cluster in a solid. Indeed, in the
presence of the phonon displacement field, ${\bf u}({\bf r},t)$,
the angle of the local rotation of the crystal lattice, $ \delta
{\bm \phi}({\bf r},t)$, and the corresponding angular velocity,
${\bf \Omega}({\bf r},t)$, are given by \cite{LL}
\begin{equation}\label{angle}
 \delta {\bm \phi}({\bf r},t) =
\frac{1}{2}\nabla \times {\bf u}({\bf r},t)\,, \qquad {\bf
\Omega}({\bf r},t) = \frac{1}{2}\nabla \times \dot{\bf u}({\bf
r},t)\,.
\end{equation}
For a transverse sound wave of frequency $f \sim 3\,$GHz and
amplitude $u_0 \sim 1\,$nm, this gives $B = \Omega/\gamma \sim
10\,$G in the rotating frame coupled to the local crystallographic
axes. Even greater local fields can be achieved with surface
acoustic waves that have been recently used in experiments on
molecular magnets \cite{Alberto1, Alberto2}.

The equivalence of the effect of high-frequency transverse
acoustic waves to the effect of high-amplitude ac magnetic field
on paramagnetic spins immediately suggests that one can try to
generate Rabi spin oscillations with the help of high-frequency
ultrasound. Rabi effect \cite{Rabi} corresponds to the oscillation
of the occupation numbers of two quantum levels in the presence of
an ac field which frequency is close to the distance between the
levels. On resonance, the frequency of Rabi oscillations is
proportional to the amplitude of the ac field. The effort to
observe Rabi oscillations between quantum states of molecular
magnets in experiments employing ac magnetic fields
\cite{Wernsdorfer,Hill,Kent,Friedman} has been going for some
time. For such experiments to succeed, the Rabi frequency must be
greater than the spin decoherence rate. This typically requires
the amplitude of the ac field to be greater than $1\,$G, which is
not easy to achieve with electromagnetic waves but, as we have
seen, is possible with surface acoustic waves. Note that the
condition of the validity of the elastic theory, $u_0 \ll \lambda$
(where $\lambda$ is the phonon wavelength) automatically provides
the condition $\Omega \ll \omega=2\pi f$, which allows one to
treat local rotations classically while treating the two-level
system with level separation $\hbar \omega$ quantum-mechanically.

\begin{figure}[h]
\unitlength1cm
\begin{picture}(7.5,4.4)
\centerline{\psfig{file=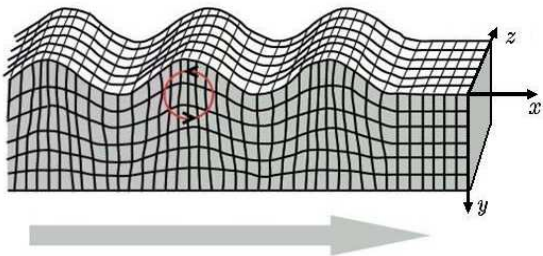,width=8.7cm}}
\end{picture}
\caption{(color online) Geometry of the problem. }\label{fig3}
\end{figure}

For certainty we consider a crystal of molecular magnets with the
anisotropy Hamiltonian
\begin{equation}\label{H-A}
\hat{\cal{H}}_{A} = -D\hat{S}_z^2 + \hat{V}\,,
\end{equation}
where $\hat{V}$ is a small term responsible for the tunnel
splitting, $\Delta$, of spin-up and spin-down states. The spin
cluster is assumed to be more rigid than its elastic environment,
so that the long-wave crystal deformations can only rotate it as a
whole but cannot change its inner structure responsible for the
parameters of the Hamiltonian $\hat{\cal{H}}_{A}$. This
approximation should apply to many molecular magnets as they
typically have a compact magnetic core inside a large unit cell of
the crystal. We choose geometry in which surface acoustic waves
are running along the $X$-axis with the solid extending towards
$y>0$, see Fig. \ref{fig3}.
Using standard formulas \cite{LL} for the displacement field in a
surface acoustic wave, one obtains
\begin{equation}\label{angle-sw}
\delta {\bm \phi}({\bf r}) = \frac{1}{2} \frac{\omega}{c_t}u_0\,
e^{-k_t y} \cos(kx - \omega t)\,{\bf e}_z \equiv
\delta\phi(x,t)\,{\bf e}_z\,,
\end{equation}
where $\omega =  c_t k\, \xi$, $k_{t} \equiv k\sqrt{1 - \xi^2}$,
$\xi$ is a real number between 0 and 1 satisfying
\begin{equation}
\xi^6 - 8\xi^4 + 8 \xi^3\left(3 - 2\frac{c_t^2}{c_l^2}\right) -
16\left(1-\frac{c_t^2}{c_l^2} \right) = 0\,,
\end{equation}
and $c_{t,l}$ are velocities of transverse and longitudinal sound.

In the presence of deformations of the crystal lattice, local
anisotropy axes defined by the crystal field are rotated by the
angle given by Eq.\ (\ref{angle-sw}). In the Hamiltonian, this
rotation is equivalent to the rotation of the operator $\hat{\bf
S}$ in the opposite direction, which can be performed by the
$(2S+1)\times(2S+1)$ matrix in the spin space \cite{CGS},
\begin{equation}\label{spin-rotation}
\hat{\bf S} \rightarrow \hat{R}^{-1}\hat{\bf S}\hat{R}, \qquad
\hat{R} = e^{i\hat{\bf S}\cdot \delta{\bm \phi}}\,.
\end{equation}
The spin Hamiltonian in the laboratory frame becomes
\begin{equation}\label{total Hamiltonian}
\hat{\mathcal{{H}}} = e^{-i\hat{\bf S}\cdot \delta{\bm
\phi}}\,\hat{\mathcal{{H}}}_A\,e^{i\hat{\bf S}\cdot \delta{\bm
\phi}}\,.
\end{equation}
In order to find the laboratory-frame wave function
$|\Psi\rangle$, it is useful to introduce the {\it lattice-frame}
wave function $|\Psi^{(lat)}\rangle$, defined through the unitary
transformation
\begin{equation}\label{psi-lat}
|\Psi^{(lat)}\rangle = e^{i\delta {\bm\phi}\cdot\hat{\bf
S}}|\Psi\rangle\,.
\end{equation}
Differentiating it on time it is easy to see that this function
satisfies Schr\"odinger equation with the lattice-frame
Hamiltonian
\begin{equation}\label{H-lat}
\hat{\mathcal{H}}^{(lat)}  =  \hat{\mathcal{H}}_A - \hbar \hat{\bf
S} \cdot {\bf \Omega}
\end{equation}
where
\begin{equation}\label{Omega}
{\bf \Omega} = \frac{\partial \delta{\bm \phi}}{\partial t} =
\frac{1}{2} \frac{\omega^2}{c_t}u_0\, e^{-k_t y} \sin(kx - \omega
t)\,{\bf e}_z\,.
\end{equation}
To this point we have not made any assumptions about the magnitude
of $\delta {\bm \phi}$, so that the equations (\ref{total
Hamiltonian}) and (\ref{H-lat}) are exact. An interesting
observation for the comparison of the effects of ultrasound and ac
magnetic field is that the Hamiltonian (\ref{H-lat}) resembles the
Hamiltonian of a particle of spin $\hat{\bf S}$ in the ac magnetic
field which amplitude scales as the square of the frequency.

We are interested in the Rabi oscillations between the two lowest
states of $\hat{\cal{H}}_A$:
\begin{equation}\label{+-}
|\phi_\pm\rangle = \frac{1}{\sqrt{2}}(|S\rangle \pm |-S\rangle
)\,,
\end{equation}
where $|\pm S\rangle$ satisfy $\hat{S}_z|\pm S\rangle = \pm S |\pm
S\rangle$. It makes sense, therefore, to project our Hamiltonian
on the $|\pm S\rangle$ states, making the problem essentially a
two-state problem. This gives
\begin{equation}\label{heff}
\hat{h}^{(lat)}_{eff} = -\frac{\Delta}{2}\, \hat{\sigma}_1 - \hbar
\omega_R\sin(kx- \omega t)\, \hat{\sigma}_3\,,
\end{equation}
where $\Delta$ is the energy distance between the ground state
$|\phi_+\rangle$ and the first excited state $|\phi_-\rangle$,
$\hat{\sigma}_1 \equiv |S\rangle\langle-S| + |-S\rangle\langle
S|$, $\hat{\sigma}_3 \equiv |S\rangle\langle S| -
|-S\rangle\langle -S|$, and
\begin{equation}
\omega_R =  \frac{1}{2 c_t}\,\omega^2\,u_0\,S\, e^{-k_t y}\,.
\end{equation}
The two-state approach will be valid if $\Delta$ and $\hbar
\omega$ are small in comparison with the distances to other spin
levels. Note that the tunnel splitting $\Delta$ originates from
the term $\hat{V}$ in $\hat{\cal{H}}_A$ that does not commute with
$\hat{S}_z$.

\begin{figure}
\unitlength1cm
\begin{picture}(20,5.4)
\centerline{\psfig{file=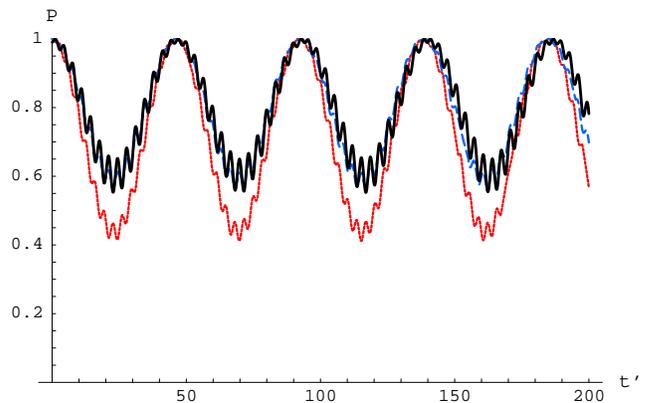,width=8.7cm}}
\end{picture}
\caption{(color online) Time dependence ($t' \equiv  t \Delta /
\hbar$) of the probability to find the spin in the state
$|\phi_{+}\rangle$ at $x = 0$, $S=10$, $\omega_R = 0.1\omega$, and
$\omega = 0.9(\Delta/\hbar)$. Dotted line (red): Numerical result
for the laboratory-frame Hamiltonian\,(\protect\ref{heffbiaxial}).
Dash line (blue): Numerical result for the lattice-frame
Hamiltonian\,(\protect\ref{heff}). Solid line (black): Analytical
result given by Eq.\ (\protect\ref{7}). }\label{fig1}
\end{figure}

It is easy to check that at $\omega \sim \Delta/\hbar$, which is
our case of interest for consideration of Rabi oscillations, the
second term in Eq.\ (\ref{heff}) can be treated as a perturbation
as long as the wavelength of the acoustic wave satisfies $\lambda
\gg Su_0$. For a not very large $S$ this condition is always
fulfilled by surface acoustic waves. The unperturbed eigenstates
of the problem are then the eigenstates of $\sigma_1$ given by
Eq.\ (\ref{+-}). Their energies are $\pm \Delta/2$. The
time-dependent perturbation produces transitions between these
states, resulting in the Rabi oscillations when $\hbar\omega
\simeq \Delta$. The standard way to obtain the evolution of the
wave function is to apply the {\it rotating wave approximation}
\cite{Rabi}. Note that the coordinates $x$ and $y$ in Eq.\
(\ref{heff}) can be viewed as parameters. Expressing the wave
function as
\begin{equation}
|\Psi(t)^{(lat)}\rangle = C_+(t) |\phi_+\rangle +
C_-(t)|\phi_-\rangle\,,
\end{equation}
and starting with $C_-(0) = 0$, $C_+(0) = 1$ at $t = 0, x=0$, one
obtains
\begin{eqnarray}\label{C}
C_-(t) & = & \frac{\omega_R}{\Omega_R}\,e^{-\frac{i}{2}\,\omega t}
\sin\left(\frac{\Omega_R t}{2} \right)\label{7} \\
C_+(t) & = & \left[\cos\left(\frac{\Omega_R t}{2} \right) + i
\frac{ \Delta/\hbar - \omega}{\Omega_R}\,\sin\left( \frac{\Omega_R
t}{2}\right) \right]e^{\frac{i}{2}\,\omega t}\,, \nonumber
\end{eqnarray}
where
\begin{equation}\label{b}
\Omega_R = \sqrt{(\Delta/\hbar - \omega)^2 + \omega_R^2}\,.
\end{equation}
Assuming that every spin was in the ground state $|\phi_+\rangle$
before the sound wave arrived, the spatial dependence of the wave
function can be obtained by making a replacement $t \rightarrow t
-kx/\omega$ in Eq.\ (\ref{C}). In the absence of spatial
derivatives in the Hamiltonian (\ref{heff}),
$|\Psi(t)^{(lat)}\rangle$ is defined up to the phase factor
$\exp[i\theta(x,y)]$ with $\theta$ being an arbitrary real
function of coordinates. From Eq.\ (\ref{psi-lat}) the wave
function of the system in the laboratory frame is $|\Psi\rangle =
e^{-i\delta {\bm\phi}\cdot{\bf S}}|\Psi^{(lat)}\rangle$. Because
of the smallness of $\delta\phi$, the dynamics of $|\Psi\rangle$
essentially coincides with the dynamics of $|\Psi^{(lat)}\rangle$
and is given by Rabi oscillations between the states
$|\phi_{\pm}\rangle$ at the frequency $\Omega_R$. This is
confirmed by numerical calculations with the lattice-frame and
laboratory-frame Hamiltonians, see Fig. \ref{fig1}.

The expectation value of the projection of the spin onto the
anisotropy axis (the Z-axis) is given by
\begin{eqnarray}\label{Sz}
& & \langle \Psi(t) | \hat{S}_z | \Psi(t) \rangle = 2S \frac{
\omega_R}{\Omega_R^2} \times \nonumber \\
&& \bigg\{\left(\omega - \frac{\Delta}{\hbar}\right) \sin(\omega t
- k x)\sin^2\left[\frac{1}{2}
\left(\Omega_R t - K_R x \right) \right] \nonumber \\
& & + \frac{1}{2}\,\Omega_R \cos(\omega t - k x) \sin
\left(\Omega_R t - K_R x \right) \bigg\}\,,
\end{eqnarray}
where $K_R = ({\Omega_R}/{\omega})k \ll k$ can be called the
``Rabi'' wave vector. Thus, the space-time Rabi oscillations of
the occupation numbers of spin states generate space-time
oscillations of the magnetization of the crystal. On resonance,
when $\omega = \Delta/\hbar$, Eq.\ (\ref{Sz}) simplifies to
\begin{equation}\label{S-res}
\langle \Psi(t) | \hat{S}_z | \Psi(t) \rangle = S \cos(kx - \omega
t)\sin\left(\omega_R t - k_R x\right)\,,
\end{equation}
with $k_R = ({\omega_R}/{\omega})k \ll k$. The condition $\omega_R
\ll \omega$ ($S u_0 \ll \lambda$) implies that the time dependence
of $\langle \hat{S}_z \rangle$ at any point in space consists of
the oscillations at frequency $\omega$ with beats of frequency
$\omega_R $. Similarly, $\langle \hat{S}_z \rangle$ at any moment
of time oscillates in space with the wave vector $k$ and exhibits
beats with the wave vector $k_R \ll k$.

Our conclusions can be checked by obtaining the full solution of
the problem in the laboratory frame in a particular case of a
biaxial symmetry, when $\hat{V}$ in Eq.\ (\ref{H-A}) equals
$E(\hat{S}_x^2 - \hat{S}_y^2)$. In this case Eq.\ (\ref{total
Hamiltonian}) reduces to
\begin{equation}\label{H-biaxial}
\hat{\mathcal{H}} = - D\hat{S}_z^2 +
\frac{E}{2}\left\{\hat{S}_+^2e^{-2i\delta\phi(x,t)} +
\hat{S}_-^2e^{2i\delta\phi(x,t)} \right\}\,,
\end{equation}
where $\hat{S}_{\pm} = \hat{S}_x \pm i\hat{S}_y$. The second term
can be treated as a perturbation provided that $E \ll D$.
\begin{figure}
\unitlength1cm
\begin{picture}(20,5.4)
\centerline{\psfig{file=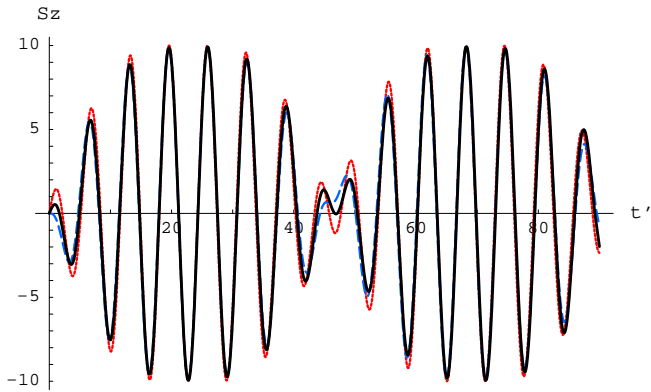,width=8.7cm}}
\end{picture}
\caption{(color online) Time dependence ($t' \equiv  t \Delta /
\hbar$) of the expectation value of the projection of the spin on
the anisotropy axis at $x = 0$, $S=10$, $\omega_R = 0.1\omega$ and
$\omega = 0.9(\Delta/\hbar)$. Dotted line (red): Numerical result
for the laboratory-frame Hamiltonian\,(\protect\ref{heffbiaxial}).
Dash line (blue): Numerical result for the lattice-frame
Hamiltonian \,(\protect\ref{heff}). Solid line (black): Analytical
result given by Eq.\ (\protect\ref{Sz}).}\label{fig2}
\end{figure}
At $\omega \ll (2S+1)D/\hbar\,$ the dynamics of the wave function
involves only a superposition of the $|\pm S \rangle$ states. As
in the lattice-frame consideration, it is then convenient to
project the Hamiltonian (\ref{H-biaxial}) onto these two states.
In order to obtain such an effective two-state Hamiltonian that
accounts for the tunnel splitting of the lowest energy states, one
must apply perturbation theory for the degenerate states $|\pm
S\rangle$ to the $S$-th order \cite{book}. This results in
\begin{equation}\label{heffbiaxial}
\hat{h}_{eff} =
-\frac{\Delta}{2}\left\{e^{2iS\delta\phi(x,t)}|S\rangle \langle
-S| + e^{-2iS\delta\phi(x,t)}|-S\rangle \langle S| \right\}.
\end{equation}
Here $\Delta = 8D {(2S)!}{\left[(S-1)!\right]^{-2}}
\left({E}/{8D}\right)^S$ is the tunnel-splitting for the biaxial
model in the absence of lattice distortions \cite{Garanin}.
Numerical solution for $\langle \hat{S}_z \rangle$ that follows
from Eq.\ (\ref{heffbiaxial}), and its comparison with the
analytical solution given by Eq.\ (\ref{Sz}), are illustrated in
Fig. \ref{fig2}. The beats discussed above are clearly seen in the
figure.

We shall now discuss conditions under which the above effects can
be observed. The first condition is that the rate of decoherence
of the spin states is lower than the frequencies involved. The
lowest of these frequencies is $\omega_R = 2\pi^2{f^2}u_0S/c_t$,
where $f$, $u_0$, and $c_t$ are the frequency, the amplitude, and
the velocity of the sound. For, e.g., $f = 3\,$GHz, $u_0 = 1\,$nm,
$S = 10$, and $c_t = 10^3$m/s, one obtains $\omega_R \sim 2\,$GHz
$\ll$ $\omega =2 \pi f \sim 20\,$GHz. If such a high value of
$\omega_R$ were to be produced by an electromagnetic wave, it
would require the ac magnetic field of amplitude $10\,$G, which is
not easy to achieve in experiment. Note, however, that the Rabi
oscillations of $\langle \hat{S}_z \rangle$ generated in a crystal
of molecular magnets by ultrasound, contrary to the Rabi
oscillations generated in a small crystal by an electromagnetic
wave, will have a pronounced wave dependence on coordinates so
that $\langle \hat{S}_z \rangle$ averaged over the wavelength of
the sound, $\lambda$, will be zero. Consequently, measurements of
the oscillations of $\langle \hat{S}_z \rangle$ should be done on
the scale that is small compared to $\lambda$.

Another restriction comes from the inevitable presence of the dc
magnetic fields that generate the Zeeman energy bias for the $|
\pm S \rangle$ states. Such fields can be of dipolar origin or
they can be any stray fields in the system. They are not likely to
affect our results qualitatively if the Zeeman energy bias is
small compared to $\hbar \omega \sim \Delta$. Note that the tunnel
splitting, $\Delta$, can be controlled by a transverse magnetic
field. Thus, the above condition translates into $B_l \leq 2 \pi
f/(\gamma S)$ for the longitudinal field $B_l$. For $f \sim
3\,$GHz one then needs $B_l \leq 100\,$G. In the case of a greater
bias field (and/or higher decoherence), higher frequencies of the
acoustic waves will be required. In principle, surface acoustic
waves of frequency as high as $100\,$GHz have been generated in
experiment \cite{Santos}. However, since $\omega_R \propto
\omega^2$, raising $\omega$ significantly may eventually violate
the condition $\omega_R \ll \omega$ under which our results were
derived.

In Conclusion, we have shown that transversal acoustic waves in
the GHz range provide spin-rotation coupling that can be used to
generate space-time Rabi oscillations in molecular magnets. When
frequency of ultrasound, $\omega$, equals the distance between
tunnel-split spin states, the magnetization on the surface of the
crystal oscillates as $\langle \hat{S}_z \rangle = S \cos(kx -
\omega t)\sin[\omega_R t - k_R x]$, where $\omega_R =
{\omega^2}u_0S/(2c_t)$ and $k_R = (\omega_R/\omega)k$, with $u_0$
and $c_t$ being the amplitude and the speed of the sound
respectively.

The authors thank Dmitry Garanin, Javier Tejada, Paulo Santos, and
Jaroslav Albert for helpful discussions. This work has been
supported by the NSF Grant No. EIA-0310517.

\end{document}